\documentclass[twocolumn,showpacs,preprintnumbers,amsmath,amssymb]{revtex4}

\usepackage{graphicx}
\usepackage{dcolumn}
\usepackage{bm}
\usepackage{amssymb}
\usepackage{graphicx}
\usepackage{amsmath}
\usepackage{xspace}

\begin{document}

\title{Polarized Raman spectroscopy of nearly-tetragonal BiFeO$_3$ thin films}
\author{M. N. Iliev}
\affiliation{Texas Center for Superconductivity and Department of
Physics, University of Houston, Texas 77204-5002, USA}
\author{M. V. Abrashev}
\affiliation{Faculty of Physics, University of Sofia, 1164 Sofia, Bulgaria}
\author{ D. Mazumdar, V. Shelke, and A.~Gupta}
\affiliation{Center for Materials for Information Technology and
Department of Chemistry, University of Alabama, Tuscaloosa, Alabama
35487, USA}

\date{\today}

\begin{abstract}
BiFeO$_3$ thin films can be epitaxially stabilized in a nearly-tetragonal phase under a high biaxial compressive strain. Here we investigate the polarized Raman spectra of constrained BiFeO$_3$ films  with tetragonal-like
(BFO-T) , rhombohedral-like (BFO-R) and multiphase (BFO-T+R)
 structure. Based on analysis of the number and symmetry of the Raman lines, we provide strong experimental evidence that the nearly-tetragonal films are monoclinic ($Cc$ symmetry) and not tetragonal $(P4mm)$. Through the Raman mapping technique we show localized coexistence of  BFO-T and BFO-R phases with the relative fraction dependent on the film thickness.
\end{abstract}


\maketitle

Constant demand for miniaturization in modern devices have provided a big stimulus for research into multi-functional materials.  BiFeO$_3$, or simply BFO, is a prototypical multiferroic material due to the simultaneous co-existence of ferroelectric, ferroelastic and anti-ferromagnetic order. It is also currently widely investigated as it offers room-temperature multi-functionality after very high polarization values were reported  in 2003 for films grown on SrTiO$_3$ (STO) substrates.\cite{wang2003} However, it can be argued that the rhombohedral $R3c$ structure of BFO-R possess significant difficulties and challenges. For example, the ferroelectric polarization of BFO-R is directed along the (111) direction giving rise to eight possible polarization orientations.  As a result, ferroelectric switching is complicated and very hard to control for any meaningful multi-functionality. Therefore, BFO phase of different symmetry, which in principle could address these issues, is desired .

This interest increased after recently it was predicted theoretically and confirmed experimentally
that the structure and properties of BiFeO$_3$ films under a large, biaxial compressive strain could deviate significantly from the bulk material into a "super-tetragonal" structure with an extremely high $c/a$ ratio.\cite{ederer2005,yun2006,bea2009,lisenkov2009,zeches2009,hatt2010} This phase, in some sense, appears incommensurable to its rhombohedral counterpart and requires a thorough investigation much along the lines that bulk BFO has received. Apart from possibly higher ferroelectric polarization values and significantly simpler switching properties, this phase is especially suitable for ultra-thin film applications where the strain effect is maximum.

Based on earlier theoretical predictions, it was anticipated that tetragonal BFO could belong to the $P4mm$~(\#99) space group \cite{ederer2005}. But careful X-ray diffraction analysis of BFO films deposited on LaAlO$_3$(LAO) and YAlO$_3$(YAO) substrates show evidence of monoclinic distortions\cite{bea2009,zeches2009}, and very recently {\it ab-initio} calculations have also indicated\cite{hatt2010} that the monoclinic $Cc$~(\#9) structure is indeed
energetically more favorable than the tetragonal $P4mm$ structure,  when the compressive strain is greater than 4\%. The three structures that could be used to describe strained BFO thin films grown on LAO substrates are shown in Fig.1.
Following the theoretical study of Hatt et al.,\cite{hatt2010} at low strains
only small differences from the structure of the bulk material are predicted and observed and
although the strained film is of lower monoclinic symmetry ($Cc$), its structure remains rhombohedral-like and to a good
approximation can be described by the $R3c$ space group. For compressive strains greater than 4\% the structure changes
dramatically and becomes tetragonal-like with $c/a\approx 1.2-1.3$. This structure can be approximated by the
$P4mm$ symmetry, but further refinement, supported by first principles calculations, leads again to $Cc$ symmetry. As
 pointed out in Refs.\cite{zeches2009,hatt2010}, the strain induced R-T transition of BiFeO$_3$ is in fact isosymmetric.
Although theoretically well grounded, there is to our knowledge no reports of direct experimental confirmation that at a
local level the structure of BTO-T films is monoclinic. Strong experimental evidence for $Cc$ structure of BFO-T films
has readily been found in the polarized Raman spectra presented and discussed below.
\begin{figure}[htbp]
\includegraphics[width=7.5cm]{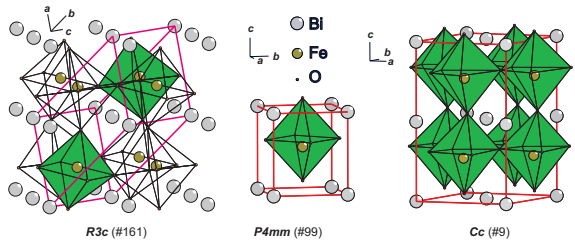}
\caption{(Color online) Elementary cells of the rhombohedral ($R3c$), tetragonal ($P4mm$) and tetragonal-like ($Cc$)
structures used to describe strained BFO/LAO films. }
\end{figure}

There are several reports on the Raman spectra of BiFeO$_3$ obtained from single crystal,\cite{fukumura2007,cazayous2007}
polycrystalline samples\cite{rout2008,yuan2007} and thin films on SrTiO$_3$ substrates.\cite{singh2005,yang2008}. These
spectra are similar with respect to the frequencies of the observed Raman lines and their
assignment is based on the rhombohedral $R3c$ structure. As an exception, Singh et al.\cite{singh2005} have
assumed that the BFO/STO thin film they studied has tetragonal $P4mm$ structure, which seems not to be correct.
Indeed, the spectra of Singh et al.\cite{singh2005} are practically identical to those from strain-free rhombohedral
BiFeO$_3$\cite{fukumura2007,cazayous2007,rout2008,yuan2007}
and, in addition, the type of substrate (SrTiO$_3$) and the film thickness (600~nm) presuppose rhombohedral-like
structure.\cite{hatt2010}

In this letter we report the polarized Raman spectra of true tetragonal-like BiFeO$_3$ (BFO-T) obtained
in strained 70~nm and 100~nm epitaxial BiFeO$_3$ films on (001)-LaAlO$_3$ substrates. The spectra are compared to those
of relaxed rhombohedral-like BiFeO$_3$ (BFO-R) obtained from 200~nm BiFO$_3$/LaAlO$_3$ and
80~nm BiFO$_3$/40~nm SrRuO$_3$/LaAlO$_3$ films. Based on analysis of the number and symmetry of the Raman lines expected for
the tetragonal $P4mm$ and monoclinic $Cc$ BFO structures, the experimental spectra provide strong evidence that the
real structure is of $Cc$ symmetry. We also show by Raman mapping of selected areas on the 100~nm~BFO/LAO and
200~nm~BFO/LAO films that BFT-T and BFO-R phases can coexist in partly relaxed BFO/LAO films.

BiFeO$_3$ thin films were deposited on single crystal LaAlO$_3$ substrates by pulsed laser deposition technique using
248~nm, KrF excimer laser with 1.5~J/cm$^2$ fluence and 10~Hz pulse repetition rate.  A ceramic target with 20\% Bi-excess was
used to compensate the Bi loss during deposition. The substrate temperature of 700$^\circ$C, oxygen pressure 100~mTorr and cool
down at 5$^\circ$C/min with 600~Torr oxygen pressure were used for all the depositions. Additionally, BFO films were deposited
 on 40 nm SrRuO$_3$(SRO) buffered LAO substrates.   X-ray diffraction measurements show BFO films (70-200 nm) deposited directly
 on LAO substrates to have an out-of-plane lattice constant value of  approximately 4.66~\AA, whereas  SRO-buffered films exhibit
 a diffraction peak near the bulk rhombohedral BFO position (pseudocubic $c \approx 3.96$~\AA). In-plane lattice parameter
 measurements through reciprocal space maps show BFO/LAO films to be highly strained to the substrate ($c \approx 3.79$~\AA).

The $R3c$ primitive cell contains two formula units. The total number of $\Gamma$-point phonon modes is 20 ($5A_1 + 5A_2 + 10E$). Of these,
13 $(4A_1 + 9E)$ are both Raman and IR active, five ($5A_2$ are silent, and two ($A_1 + E$) are acoustical modes.
The primitive cell of the $P4mm$ structure (Fig.1) contains one formula unit. The total number of $\Gamma$-point phonon modes
is 10 ($4A_1+B_1+5E)$. Of these, seven $(3A_1 +4E)$ are both Raman and IR active, one $(B_1)$ is only Raman active and
two $(A_1+E)$ are acoustic modes. The elementary cell of $Cc$ structure (Fig.1) is base-centered and contains four formula
units. All atoms are non-centrosymmetrical  $(4a)$ positions. The number of $\Gamma$-point phonon modes is 30 $(15A'+15A'')$. Three of
them $(A'+2A'')$ are acoustic modes. $14A'+13A''$ modes are both Raman and infrared active.
As the experimentally obtained out-of-plane lattice parameter of the BFO-T films is larger than the in-plane parameters,
it is plausible to accept that the surface of the film is parallel to the corresponding (001) planes. We note here that the
[100]$_t$ and [010]$_t$ directions of $P4mm$ structure, which are parallel to [100]$_c$ and [010]$_c$ quasicubic directions
of the LAO substrate, become [110]$_m$ and [-110]$_m$ directions for the $Cc$ structure. Therefore, the coordinate systems
of the Raman tensors for the $P4mm$ and $Cc$ differ by 45 degree rotation around the $c$-axis.

The scattering intensity of a phonon mode of given symmetry is proportional to
$(\vec{e}_i \mathbf{R} \vec{e}_s)^2$ ,
where $\vec{e}_i$ and $\vec{e}_s$ are unit vectors parallel, respectively, to the polarization of the incident
and scattered radiation. The Raman tensors for modes of different symmetry of the $R3c$, $P4mm$ and $Cc$ structures
have the form:
\begin{center}

 \begin{tabular}{ccc}
\ & $A_1$ & $E$\\

$R3c$ \ $\Rightarrow$& $\left[
\begin{array}{ccc}
 a & \ & \ \\
 \ & a & \ \\
 \ & \ & b
 \end{array} \right]$&
$\left[
\begin{array}{ccc}
 c & \ & d \\
 \ & -c & \ \\
 d & \ & \
 \end{array}\right],
 \left[ \begin{array}{ccc}
\ & -c & \ \\
-c & \ & d \\
 \ & d & \
 \end{array}\right]$\\

 \ &\ & \ \\

 \end{tabular}

\begin{tabular}{cccc}
\ & $A_1$ & $B_1$ & $E$ \\

$P4mm$ \ $\Rightarrow$ & $\left[
\begin{array}{ccc}
 a & \  & \ \\
 \ & a & \ \\
 \ & \ & b
 \end{array} \right]$&
$\left[
\begin{array}{ccc}
 c & \  & \ \\
 \ & -c & \ \\
 \ & \ & \
 \end{array} \right]$&
$\left[
\begin{array}{ccc}
 \ & \ & e \\
 \ & \ & \ \\
 e & \ & \
 \end{array}\right],
 \left[ \begin{array}{ccc}
\ & \ & \ \\
\ & \ & e \\
 \ & e & \
 \end{array}\right]$\\
 \end{tabular}

\

and

\

 \begin{tabular}{ccc}
\ & $A'$ & $A''$\\

$Cc$ \ $\Rightarrow$& $\left[
\begin{array}{ccc}
 a & \  & d \\
 \ & b & \ \\
 d & \ & c
 \end{array} \right]$&
$\left[
\begin{array}{ccc}
 \ & e & \ \\
 e & \ & f \\
 \ & f & \
 \end{array} \right]$\\
 \end{tabular}
 \end{center}

For backward scattering from the (001) surface $\vec{e}_i$ and $\vec{e}_s$ have no $z$-component and
the intensity of the $E$ modes (for $P4mm$) will {\it a priori} be zero. One therefore expects
observation in the experimental Raman spectra of BFO-T of only four  $(3A_1+B_1)$ Raman lines  in the case of $P4mm$ and much
higher number of lines $(14A'+13A'')$ for the $Cc$ structure. The polarization selection rules for these Raman mode symmetries in
all available exact scattering configurations from the tetragonal $(001)_t$ or monoclinic $(001)_m$ surfaces are given
in Table I. For the $Cc$ structure the intensities are averaged over the expected two twin variants with interchangeable $a$ and $b$
parameters (four twin variants if accounting for the polarization direction). Similarly, for the BFO-R films four twin variants
(eight if accounting for the polarization direction) with orientation of the rhombohedral [111]$_r$ direction along any
of the four quasicubic directions ${111}_c$ of the LAO substrate
may coexist. The polarization selection rules for the $R3c$ and $Cc$ structures, given in Table~I, are the expected
averaged Raman intensities with $\vec{e}_i$ and $\vec{e}_s$ along the cubic $(x$, $y$, $x'$, and $y'$ directions, under the
assumption that the twin variants occupy equal parts of the scattering volume.

\begin{table*}
\caption{Polarization selection rules for the phonons of non-zero intensity with backward scattering from the $(001)_c$
plane of $R3c$, $(001)_t = (001)_c$ plane of $P4mm$, or $(001)_m$ plane of $Cc$ structures. The intensities for the $R3c$
and $Cc$ modes are averaged over the expected twin variants.}
\begin{tabular}{|ccccccc|}

\hline
     &$(xx)_c$&$(yy)_c$&$(xy)_c$&$(x'x')_c$&$(y'y')_c$&$(x'y')_c$\\
Mode &$(xx)_t$&$(yy)_t$&$(xy)_t$&$(x'x')_t$&$(y'y')_t$&$(x'y')_t$\\
     &$(x'x')_m$&$(y'y')_m$&$(x'y')_m$&$(xx)_m$&$(yy)_m$&$(xy)_m$\\
 \hline
 & & & & & & \\
$A_1 (R3c)$&$\frac{1}{9}(2a+b)^2$&$\frac{1}{9}(2a+b)^2$&$\frac{1}{9}(a-b)^2 \approx 0$&$\frac{1}{2}[a^2 + \frac{1}{9}(a+2b)^2$&$\frac{1}{2}[a^2 + \frac{1}{9}(a+2b)^2$&$0$\\
$E(R3c)$&$\frac{4}{9}(-c+\sqrt{2}d)^2$&$\frac{4}{9}(-c+\sqrt{2}d)^2$&$\frac{1}{9}(2c+\sqrt{2}d)^2$&$\frac{1}{2}[c^2 + \frac{1}{9}(c+2\sqrt{2}d)^2$&
$\frac{1}{2}[c^2 + \frac{1}{9}(c+2\sqrt{2}d)^2$&$\frac{1}{3}(-c+\sqrt{2}d)^2$ \\

 & & & & & & \\
$A_1 (4mm)$&$a^2$&$a^2$&$0$&$a^2$&$a^2$&$0$\\
$B_1(4mm)$&$c^2$&$c^2$&$0$&$0$&$0$&$c^2$\\
 & & & & & & \\
$A' (Cc)$&\ $ \frac{1}{4}(a+b)^2$ \ & \ $\frac{1}{4}(a+b)^2$ \ &$ \ \frac{1}{4}(a-b)^2 \approx 0$ \ & \ $ \frac{1}{4}(a+b)^2$ \ & \ $ \frac{1}{4}(a+b)^2$ \ & \ $0$ \ \\
$A'' (Cc)$&\ $ e^2$ \ & \ $e^2$ \ &$0$ \ & \ $0$ \ &$ 0$ \ & \ $e^2$ \ \\
 & & & & & & \\
  \hline
\end{tabular}
\end{table*}

Figure 2  illustrates on the example of 70 nm BFO/LAO film how the spectra of BFO-T were obtained. Due to the small film
thickness, the original spectra are a superposition (BFO+LAO) of Raman signals from the film (BFO) and the substrate (LAO).
The LAO spectra were measured separately with the same scattering configuration from film-free substrate surface and then
subtracted from the original spectra using GRAMS AI software.
\begin{figure}[htbp]
\includegraphics[width=7cm]{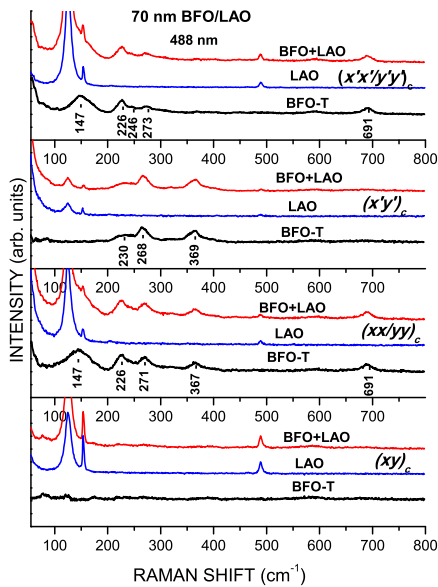}
\caption{(Color online) Polarized Raman spectra of BFO/LAO, LAO and extracted spectra of pure BFO-T as obtained
with 488 nm excitation from a 100~nm BFO/LAO film.}
\end{figure}

In Figure 3 are compared the spectra of BFO-T, obtained with 515~nm and 488~nm excitation from two strained BFO-T films, with
the corresponding spectra of relaxed BFO-R, obtained with 488 nm excitation. As it follows from Table~I, the $A_1$ modes of BRO-R
structures are much stronger with parallel $(xx/yy)$ and $(x'x'/y'y')$ than with crossed $(xy)$ and $(x'y')$ scattering configurations.
This allows to identify unambiguously the peaks at 77, 142, 176, and 221~cm$^{-1}$ in the BFO-R spectra as corresponding to the
$A_1$ modes. The assignment of the 77~cm$^{-1}$ line to an $A_1$ mode differs from that proposed by Cazayous et al.,\cite{cazayous2007}
but as a whole the BFO-R spectra are similar to those reported for single crystal and bulk BFO.\cite{fukumura2007,cazayous2007,rout2008,yuan2007}.
The weaker peaks at 279, 359, 369, 473, 530, and 615~cm$^{-1}$ can be assigned to $E$ modes.
\begin{figure}[htbp]
\includegraphics[width=7cm]{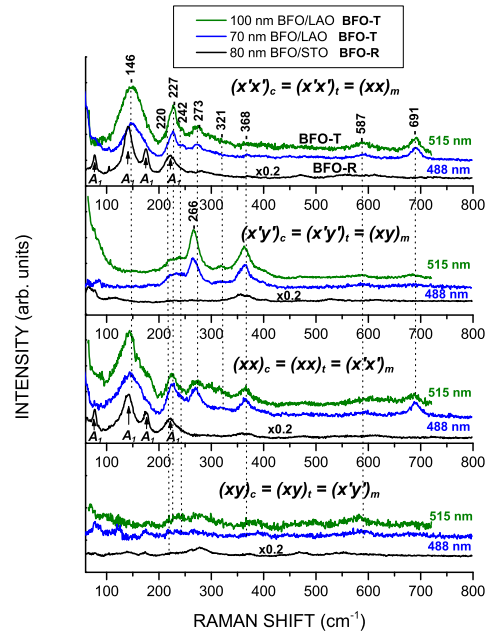}
\caption{(Color online) Comparison of the spectra of BFO-T obtained from 70 nm BFO/LAO and 100 nm BFO/LAO with the spectra
of BFO-R obtained from 80 nm BFO/STO/LAO thin films. The spectra with 488 nm excitation were obtained at the same experimental conditions.
The spectra of BFO-R are scaled by factor of 0.2.}
\end{figure}

The Raman spectra of BFO-T are well reproduced in all scattering configurations. The large number of Raman lines is consistent with
monoclinically distorted tetragonal-like $Cc$ structure, thus ruling out the simple $P4mm$ structure. Indeed, for the $P4mm$ structure one
expects only three $A_1$ lines in the $(x'x')_c$ and one $B_1$ line in the $(x'y')_c$ spectra,
whereas the number of experimentally observed lines is much higher. For the same scattering configurations 14~$A'$ and
13~$A''$ modes are allowed, respectively, for the $Cc$ structure. We can therefore assign the Raman lines at 146, 227, 273, 587, and 691~cm$^{-1}$,
pronounced in the $(xx)_c$~($A'+A''$) and $(x'x')_c$~($A')$ spectra, to modes of $A'$ symmetry and the lines at 220, 242, 266, and 368~cm$^{-1}$,
seen in the $(xx)_c$~($A'+A''$) and $(x'y')_c$~($A'')$ spectra, to modes of $A''$ symmetry

With increasing BFO/LAO film thickness the rhombohedral BFO-R phase appears as secondary phase in the 100~nm~BFO/LAO film and dominant
phase in the 200~nm~BFO/LAO film. This is illustrated for $45 \times 45\  \mu m^2$ area on the surface of 100~nm BFO/LAO thin film, which has been Raman mapped with 1.5~$\mu$m step comparing the $(xx)_c$ Raman spectra. As it follows from Fig.4, the studied area is characterized by three
types of spectra: BFO-R1, BFO-R2, and BFO-T. The Raman line frequencies of BFO-R1 and BFO-R2 correspond to the rhombohedral phase and
the spectra differ by only relative line intensities. This can be explained by assuming that the BFO-R1 and BFO-R2 spectra are obtained
from twin variants of BFO-R with different orientation with respect to $\vec{e}_i$ and $\vec{e}_s$. One can therefore conclude, that the
twin variants of the BFO-R phase are of micrometer size. As to the twin variants of the BFO-T phase, their Raman spectra should be
practically undistinguishable, as from structural considerations one expects the relation $a \approx b$ for the components
of the $A'$ Raman tensor. The observation of the R-phase is probably due to the release of the substrate strain effect with increasing thickness.\cite{zeches2009}

\begin{figure}[htbp]
\includegraphics[width=5.5cm]{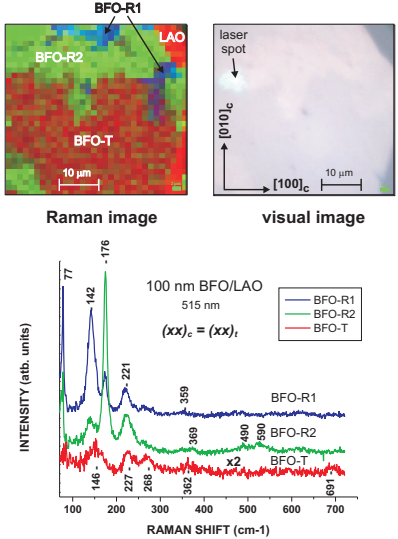}
\caption{(Color online) Raman mapping (upper left) and visual image (upper right)
of the same  $45 \times 45\  \mu m^2$ area on the surface of 100~nm BFO/LAO thin film.
The blue and green colors indicate presence of BFO-R twin variants of different orientation.
The red colored area corresponds to the BFO-T phase. In the bottom panel are
shown the corresponding Raman spectra obtained with $(xx)_c$ scattering configuration
with 515~nm excitation.}
\end{figure}

In conclusion, polarized Raman spectroscopy was used to study tetragonally strained BFO-T films on LAO substrates and relaxed rhombohedral BFO-R films
on LAO substrates with SRO buffer layers. We provide strong experimental evidence that the tetragonal-like structure is of monoclinic $Cc$ symmetry and
the simple tetragonal $P4mm$ structure has to be ruled out. Five of the fourteen $A'$ modes and four of the thirteen $A''$ modes
expected for the $Cc$ structure
have been identified This is consistent with the results of recent first principles calculations\cite{hatt2010}. It is also demonstrated very clearly that BFO-R and BFO-T coexist at the micrometer scale with the R-phase gradually becoming the dominant phase with increasing thickness.
\acknowledgments
This work was support in part by the State of Texas through the
Texas Center for Superconductivity at the University of Houston
and by NSF MRSEC (Grant No.DMR-0213985).

\end{document}